\documentstyle[12pt,epsfig]{article}
 1
 1
 1

\begin{document}
\bibliographystyle{unsrt}
% This defines greater than, less than, greater or less, less or greater 
% approx. symbols
%
\def\lta{\;\raisebox{-.5ex}{\rlap{$\sim$}} \raisebox{.5ex}{$<$}\;}
\def\gta{\;\raisebox{-.5ex}{\rlap{$\sim$}} \raisebox{.5ex}{$>$}\;}
\def\grle{\;\raisebox{-.5ex}{\rlap{$<$}}    \raisebox{.5ex}{$>$}\;}
\def\legr{\;\raisebox{-.5ex}{\rlap{$>$}}    \raisebox{.5ex}{$<$}\;}

% This defines exponents and indices in text mode 
%
\def\r#1{\ignorespaces $^{\rm #1}$} 
\def\l#1{\ignorespaces $_{\rm #1}$} 

\newcommand{\ra}{\rightarrow}
\newcommand{\permille}{$^0 \!\!\!\: / \! _{00}\;\;$}
\newcommand{\dd}{{\rm d}}
\newcommand{\oal}{{\cal O}(\alpha)}%
\newcommand{\su}{$ SU(2) \times U(1)\,$}
 
\newcommand{\eps}{\epsilon}
\newcommand{\lra}{\leftrightarrow}
\newcommand{\tr}{{\rm Tr}}
 
\newcommand{\ie}{{\em i.e.}}
\newcommand{\cm}{{{\cal M}}}
\newcommand{\cl}{{{\cal L}}}
\def\nn{\noindent}
%******
\newcommand{\be}{\begin{equation}}
\newcommand{\ee}{\end{equation}}
\newcommand{\ba}{\begin{eqnarray}}
\newcommand{\ea}{\end{eqnarray}}
%******
\newcommand{\nl}{\nonumber \\}
\newcommand{\eqn}[1]{Eq.(\ref{#1})}
\newcommand{\ibidem}{{\it ibidem\/},}
\newcommand{\into}{\;\;\to\;\;}
\newcommand{\wws}[2]{\langle #1 #2\rangle^{\star}}
\newcommand{\cjg}{^{\star}}
\newcommand{\lgn}[1]{\log\left(#1\right)}
\newcommand{\p}[1]{{\scriptstyle{\,(#1)}}}
%*******************
\newcommand{\mw}{M_{W}}
\newcommand{\mbb}{m_{b \bar b}}
\newcommand{\mff}{m_{f \bar f}}
\newcommand{\seff}{s_{eff}}
\newcommand{\mh}{m_{H}}
\newcommand{\mz}{M_{Z}}
\newcommand{\mt}{m_{top}}
\newcommand{\epem}{e^{+} e^{-}}
\newcommand{\eeww}{e^{+} e^{-}\to W^+W^-}
\newcommand{\gaga}{\gamma \gamma }
\newcommand{\si}{\sigma}
\newcommand{\sit}{\sigma_{tot}}
\newcommand{\sitww}{\sigma_{tot}(WW)}
\newcommand{\gev}{GeV}
\newcommand{\sqs}{\sqrt{s}}
\newcommand{\sqsp}{\sqrt{s'}}
%**********************
\begin{titlepage}
\rightline{ROME1-1173/97}
\rightline{May 1997}
\rightline{hep-ph/9705379}
\vskip 22pt 
\bigskip
\vspace*{2cm}
\begin{center}
{\Large \bf Physics at LEP2 
  }
\end{center}
\bigskip

\begin{center}
{\large 
B.~Mele
} \\

\medskip

INFN, Sezione di Roma 1 and Rome University ``La Sapienza", Italy
\end{center}
\bigskip
\bigskip
\begin{center}
{\bf Abstract} \\
\end{center}
{\small 
The main physics topics of interest at LEP2, the CERN  electron-positron 
collider with center-of-mass energy in the range (161-192) GeV, are reviewed.
Progresses in both the precision tests of the standard model
and the field of discovery physics attainable at this machine
are discussed. A few results of the analysis of data collected in the 1996 
first runs of LEP2 are also presented.
}

\vskip 22pt 
\vfil
\noindent
e-mail: mele@roma1.infn.it \\
Lectures given at the First School of Field Theory and Gravitation,
14-19 April, 1997,  Vitoria (Brasil). 

%\vspace*{1cm}
%\leftline{FNT/T-95/32}
\end{titlepage}
%%%%%%%%%%%%%%%%%%%%%%%%%%%%%%%%%%%%%%%%%%%%%%%%%%%%%%%%
%%%%%%%%%%  End of the cover page  %%%%%%%%%%%%%%%%%%%%%
%%%%%%%%%%%%%%%%%%%%%%%%%%%%%%%%%%%%%%%%%%%%%%%%%%%%%%%%
\noindent
\section{Introduction}
\vskip 10pt

After a short run at intermediate energy in the 1995 fall 
(the so called LEP1.5 phase),
LEP2, the second phase of LEP, the CERN Large Electron-Positron collider
with a centre-of-mass energy in the range
$\sqs\simeq (161-192)$GeV, started operating. 
This follows the excellent performance
of LEP1 during the years 1990-95.
LEP1 collected about 145 pb$^{-1}$ of integrated luminosity,
working most of the time on the $Z$ vector-boson peak, at
$\sqs\simeq \mz$. In particular,  about 16 million of $Z$ 
events have been analyzed, mostly associated to the two-fermion processes:
$$
e^+e^-\to Z^0 \to \ell^+  \ell^-, \nu \bar \nu , q \bar q,
$$
where $\ell^{\pm}$ can be an electron, muon or tau lepton, and
$\nu$ and $q$ are neutrinos and  quarks, respectively.
 The fine structure of the standard model was tested
at an unprecedented (and unexpected before the LEP performance) 
level at LEP1.
A summary of the most important results on precision measurements
at LEP1 can be found in Table 1, where the numbers quoted
at the 1997 Moriond Conference \cite{mori} are presented.
Apart from the comparison between the measured LEP1 value and the SM 
expectation, for each quantity, the {\it pull} is shown,
where the pull is defined as the ratio of the difference
between the measured and the predicted values 
over the corresponding experimental errors.
One can note that, on the one hand, the experimental relative error
is very small in several cases (e.g., only $2\cdot 10^{-5}$ 
on the $Z$ mass 
and about 1 \permille on the $Z$ width $\Gamma_Z$, the hadronic cross section
$\sigma_h$
and the  $\sin^2\theta_{eff}$ electroweak parameter).
Furthermore, the pull is quite low (in absolute value)
and points to an excellent agreement between the experimental results and
the SM predictions in any single detail of the theory.
As a result, LEP1 provided a beautiful and highly nontrivial
consistency check of the theory.
On the other hand, on the same basis, one can even put strong limitations 
on  models predicting physics beyond the standard model.

Compared to LEP1, the LEP2 physics potential presents
quite different features.
First of all, LEP2 works far away the $Z$ peak, and beyond the $\eeww$
threshold (i.e., at $\sqs\gta 2\mw$). 
The cross sections for interesting processes at LEP2
are in general a few 10-pb's, to be compared with the typical LEP1
cross section of $\sim 10$ nb. With the expected LEP2 final integrated 
luminosity of 500pb$^{-1}$, this corresponds to a few 10$^4$ events
collected at the end of the operation of LEP2, 
to be compared with the 16 million of $Z$ events analyzed at LEP1.
Hence, the typical LEP2 statistical  error 
($\sim 1/\sqrt{N}$) is of the order of 
1 \%, instead of the corresponding $\sim 1$ \permille at LEP1.
LEP2 is not expected to be, at least as much as LEP1 has been,
a precision measurement tool, 
 although in a couple of important cases, 
that we will consider below, LEP2 will produce new precision results.

The basic feature of LEP2 is instead that it enters for the first time into 
a new energy regime for electron-positron collisions. This implies the 
possibility of exploring direct new physics signals in the electroweak sector
of the theory (the most difficult to test at hadron colliders), 
with typical masses for the new degrees of freedom
as large as  about $\sqs/2 \; \; (i.e., \lta 100 GeV$), and sometimes 
even larger than that.

In these lectures, I will first present the standard model 
expectations for the physics at LEP2. This is, in a sense, a minimal
scenario that we are confident to observe and study at the LEP2 experiments.
I will start by the two-fermion processes and the radiative-return 
phenomenon. Then, the $WW$ production will be discussed both 
as a test of possible 
anomalies in the trilinear vector-boson ($WW\gamma$ and $WWZ$) vertices,
and as a means to determine $\mw$ precisely. 
More generally, the whole class of four-fermion processes 
(i.e., $ee\to 4f$), whose accurate
knowledge is crucial at LEP2, will be analyzed.
Possible checks of 
the quantum chromodynamics (QCD) at the energy scale $Q\sim \sqs \sim 200 GeV$
will be also stressed.
Then, I will present a few major topics on discovery physics.
In particular, I will discuss the LEP2 potential in  
searching for the Higgs boson. Expected signals from 
a low-energy supersymmetric extension of the standard model
will be also discussed.

When available, I will combine the theoretical predictions with the 
experimental results and analysis from the first 1996 runs of LEP2.
In 1996, LEP2 collected about 11 pb$^{-1}$ of integrated luminosity
on the $\eeww$ threshold (i.e., at $\sqs\simeq 2\mw \simeq 161$GeV), 
then about 0.4 pb$^{-1}$ and 1 pb$^{-1}$ during  
two short runs at $\sqs\simeq 164$ and 
170 GeV, respectively, and, finally, 
about 11 pb$^{-1}$ at $\sqs\simeq 172$GeV.
A year before, in 1995, LEP had also run at intermediate energy,
$\sqs\simeq$130-136 GeV, with a total luminosity of about  6 pb$^{-1}$.
The machine is now expected to run at higher energy ($\sqs\simeq 188$GeV),
starting from  June '97, and to reach the maximum energy of 
$\sqs\simeq 192$GeV by 1998. The expected integrated luminosity
is about 100-150 pb$^{-1}$/year per experiment,
in the three years '97,'98 and '99.

Further material, details and useful references
on the topics treated in these lectures can be found in \cite{lepd}.
\noindent
\section{Standard model physics}
\vskip 10pt

In figure 1 \cite{lepd},  
the cross sections for the main standard model processes
relevant at LEP2 are reported (in pb)
versus the center-of-mass energy.
The final state associated to each curve is shown in the figure.
The total number of the events expected at LEP2 for  the corresponding
channel, 
after  the complete period of  operation, can be
straightforwardly obtained by multiplying $\sigma[(161-192) GeV]$ in pb 
by 500, corresponding to an integrated luminosity of 500 pb$^{-1}$.
One can observe that
there are many competing processes at the LEP2 center-of-mass energies, 
contrary to the  LEP1 case.
Getting further from the $Z$ peak, the 
fermion-pair production (represented for its hadronic component by 
the $\Sigma q \bar q$ curve) drops as $1/s$, while the $WW$ production
has  a comparable cross section for $\sqs\gta 170 GeV$.
Other very important channels are the $\gaga$ reactions, 
where the two fermions are produced by the interaction of 
two almost-real photons radiated by forward-scattered electrons and positrons
(see figure 2). This class of processes (although of higher order in the 
electroweak coupling) has a very large cross section, which
nevertheless decreases when cutting on the low invariant-mass range 
of the produced fermion-pair.
For instance, the cross section shown in figure 1 
for $\epem \to \epem \tau^+ \tau^-$ corresponds to cutting away
all the events with $m_{(\tau^+ \tau^-)}<20$GeV, and is still larger 
than 10 pb (i.e., it will give more than 5000 events) at LEP2.
The associated production of one photon plus a $Z$ boson and of two photons
at large angles (e.g., $|\cos\theta_{e\gamma}|<0.9$ in figure 1)
has  also relevant rates.
Lower rates characterize the production of one $\gamma,Z,W$ plus
two leptons. Furthermore, above the $ZZ$ threshold, 
the two-$Z$ production has a cross section larger than 1 pb.
%******************************************
\subsection{Z radiative return}
The curve named $\Sigma q \bar q$(ISR) in figure 1, 
corresponds to the inclusion of the
Initial-State-Radiation effects in the two-quark production.
The ISR gives  the main contribution to the total QED corrections.
Although the  higher-order corrections should in general
moderately alter the tree-level values, in this case one observes an
increase of about a factor 4 in the total rate at LEP2 energies.
This can be explained by what is usually called the {\it $Z$ radiative return}.
It corresponds to the dominance in the radiative corrections of the
$\epem\to f\bar f$ channel at $\sqs>\mz$ of a kinematical configuration
where the invariant mass of the fermion pair $\mff$ is still 
about $\mz$ and a real photon is radiated with energy
$$
E_{\gamma}^0=\frac{s-\mff^2}{2\sqs}\simeq \frac{s-\mz^2}{2\sqs}.
$$
The effect arises from the Breit-Wigner behavior 
of the two-fermion cross section 
for $s$ around the $Z$ peak, which makes the rate for 
off-shell fermion pair production, and in particular for $\mff>\mz$,
much smaller than on the $Z$ peak, and
{\it of the same order in the electroweak coupling} 
of the process of a $real$-$Z$ production, 
accompanied by a real photon, that takes away the residual
energy. The real $Z$ subsequently decays into a $f \bar f$ pair with 
$\mff\simeq\mz<\sqs$.
The consequent photon energy spectrum in the ISR corrected distribution
shows a peak around the value $E_{\gamma}^0$.
Correspondingly, the corrected $\mff (\equiv \sqsp)$ 
distribution shows a reduced peak at $\sqsp \simeq \sqs$, while the bulk of
the cross section is concentrated around $\sqsp \sim \mz$ 
(see figure 3, for the quark pair
production $\epem\to q\bar q \gamma$ \cite{ltce}).
In figure 4 \cite{prec}, the continuous lines show the theoretical predictions
for the almost inclusive measurement with $\mff>0.1 \sqs$ (solid line)
and the ``real" LEP2 events 
(i.e., dropping off $Z$-radiative-return events)
with $\mff>0.85 \sqs$ (dashed line).
In the same figure, the results from LEP1, LEP1.5 and LEP2
(combining the four experiment data) are reported.
The relative difference between the data and the SM predictions 
is reported in the
lower plot of figure 4, that shows good agreement for both hadronic and
leptonic data.

\subsection{W pair production}
LEP2 was mainly designed to work above the $WW$ threshold. This allows to study
for the first time in $\epem$ collisions the direct production of $W$
vector bosons. At tree level, the $W$ pair production occurs via the 
three graphs in figure 5, i.e., through either  $t$-channel neutrino exchange
or $s$-channel $\gamma$ and $Z$ exchange, the latter involving non-abelian
three-vector-boson vertices. These two modes give 
separately divergent contributions to the total cross sections,
 violating unitarity at high energies with terms of the order $s/\mw^4$
(see the dashed lines in figure 6 \cite{denn}).
An accurate gauge cancellation between the two terms ensures
the good high-energy
behavior of the cross section, with $\sit\sim  \log{\frac{s}{\mw^2}}/s$
(solid line in figure 6). \\
\noindent
As a consequence, the $WW$ cross section is very sensitive to any deviation
from the exact gauge cancellation, and its measurement gives  a good test
of the non-abelian sector of the standard model.
In particular, limits on possible anomalies in the $\gamma WW$ and $ZWW$ 
vertices can be set at LEP2, complementing and improving the
corresponding limits worked out in $pp$ collisions at the Tevatron.
Dedicated studies have shown that limits 
on gauge anomalous coupling  attainable at LEP2
can be made stronger, if, apart from the 
$WW$ total cross section information, one exploits the full
angular-distributions information on both the $W$ production angles and 
the $W\to \ell\nu,jj$ decay angles. For instance, in figure 7
\cite{lepi}, the contours of limits
on the two parameters $\delta\kappa_{\gamma}$ and $\lambda_{\gamma}$,
parametrizing the possible C and P conserving $\gamma WW$ anomalies,
are shown, for different cases of available angular information,
at $\sqs=176$ and 190 GeV [outermost contour uses only the $W$ production
angle, while the innermost one uses all the available
(production and decay) angular data from the 
semileptonic $\ell\nu jj$ state]. Here, one assumes the final integrated
luminosity of the machine. \\
\noindent
Present LEP2 data on the $W$ anomalous couplings correspond to a luminosity 
still too low  to improve the constraints on the $W$ vertices 
anomalies obtained at other machines.

The second basic aim of the study of the $WW$ production at LEP2,
is the precise direct determination of the $W$ mass.
Such a measurement complements the precision measurement
results of LEP1.
Indeed, from LEP1, one gets only an indirect determination of $\mw$
through the relation
$$
G_{\mu}=\frac{\alpha \pi}{\sqrt 2 \mw^2(1-\mw^2/\mz^2)}
        \frac{1}{1-\Delta r}
$$        
where the Fermi Constant $G_{\mu}$ is accurately known from muon decay,
and $\Delta r$ is 0 at tree level, while is linearly 
dependent on  $\mt^2$ and $\log{\mh}$, when the dominant 
loop corrections are included. Hence, at LEP1, the $\mw$ determination depends
on the values of $\mt$ and $\mh$. At LEP2, the direct measurement
of $\mw$ allows instead a consistency check of the standard model
through the relation above. On the other hand, by inserting the $\mt$
value measured at TeVatron, one can get through the same relation
some indication on the $\mh$ value.
For instance, in Table 2 \cite{lepd}, for two different assumptions 
on the final indetermination
 on $\mw$ (25 or 50 MeV) and $\mt$, the estimated error on $\mh$
for different ``central" (in the logarithmic scale) values for $\mh$
are reported. One can see that, especially for low values of $\mh$,
assuming a very good resolution on $\mw$, such as 25 MeV, is equivalent 
to an indirect measurement of $\mh$ with a precision of a few tenths of GeV's.

On the experimental side, there are two main methods to determine
$\mw$ at LEP2. 
The first consists in measuring the total cross section for $\eeww$
on the threshold (i.e., at $\sqs \simeq 161$ GeV). 
This exploits the fact that, although the cross section is not maximal 
at that energy ($\sitww_{161} \simeq 3.7$ pb), 
the statistical experimental error is strongly reduced by its inverse 
dependence on $d\sigma/d\mw$, which is maximal at the threshold.
The second method consists in directly reconstructing the
resonance $\mw$ peak in the invariant mass distributions  of the
$W$ decay products, by using either the semileptonic mode 
($WW \to \ell \nu jj$) or the hadronic mode ($WW \to jjjj$). 
These measurements are carried out above the threshold 
(in particular, for $\sqs \gta 170$ GeV),
where the larger cross section ($\sitww_{170} \simeq 12$ pb) 
reduces the statistical
error, that scales as the ratio of the $W$ width over the square
root of the total number of the events.
In figure 8 \cite{prec}, the final results for the $\mw$ measurement 
on threshold
are reported. The rather large final error (about 200 MeV) suffers from
the reduced luminosity collected at $\sqs=161$ GeV (about 11 pb$^{-1}$)
with respect to the planned value  (50 pb$^{-1}$), that would have given
$\Delta \mw \sim 100$ MeV.
In figure 9 \cite{prec}, the results based on 
the direct recostruction method at $\sqs=172$GeV are shown. 
This kind of measurement will be much improved
by the further runs of LEP2. The present error on the directly 
reconstructed $W$ mass is anyway already comparable (about 190 MeV)
to the measurement on threshold.
By averaging the two  measurements, one obtains the preliminary
LEP2 result $\mw=80.38 \pm 0.14$ GeV.
One can then combine this result with the present $\mw$ measurement
at the TeVatron ($\mw=80.37 \pm 0.10$ GeV), and 
obtain the present world average
$$
\mw=80.37 \pm 0.08 \; \; \gev.
$$
One can notice that, although still in a preliminary stage, LEP2
already contributes in getting a precision in the direct $\mw$ determination
as good as 80 MeV.

LEP2 will also be able to improve the present determinations of the $W$
branching fractions in the different decay channels. Although, 
with the present statistics, the LEP2 data allow the BR($W$) measurements 
only with a 
rather low accuracy with respect to the corresponding TeVatron 
determinations, the LEP2
complete run will quite improve this sector of the $W$ physics, too.

\subsection{Four-Fermion Processes}
At  LEP2 centre-of-mass energies, four-fermion final states are 
produced with large 
cross sections. These are not only due to real $WW$ and $ZZ$
pair production with subsequent decays
$W\ra \bar{f} f'$ and $Z\ra \bar{f} f$, but arise
from  several production mechanisms, each  giving 
sizeable  contributions
to the four-fermion cross section in specific configurations of the 
final-particle phase space.
In figure~10, all the possible classes of four-fermion production
diagrams are shown. The largest total cross sections arise from
the {\it multiperipheral}  diagrams. Here, 
two quasi-real photons are 
exchanged in the $t$-channel, giving rise to forward (and
undetected) electrons/positrons plus a $\bar{f} f$ pair with 
a non-resonant structure (the so-called ``two-photon'' processes).
For instance, one has $\sigma(\epem\ra \epem \tau^+\tau^-)\sim 10^2$ pb for
$M_{\tau\tau}>10$GeV. On the other hand, although interesting for QCD 
studies and as a main background for missing energy/momentum events,
these classes of processes do not sizably contribute
to final states that are of interest for the studies of $W,Z$ and Higgs
boson production.  In the latter case, the main contributions
come from the double-resonant diagrams ({\it conversion} and
{\it nonabelian-annihilation} diagrams in figure~10).
Also single-resonant processes (proceeding through
{\it abelian-annihilation, bremsstrahlung, fusion} and
single-resonant {\it conversion}
graphs) can give an important contribution
to vector-boson physics, when the invariant mass
constraint on one of the final fermion pairs is relaxed.
A particular example is given by the single $W,Z$
production, $\epem \ra e\nu W\ra e\nu f f'$ and
$\epem \ra e e Z \ra ee ff$. In this case, 
most of the cross section is due to single-resonant
{\it bremsstrahlung} and {\it fusion} diagrams,
where an almost real photon is exchanged in the $t$-channel
and one final electron escapes detection.
In a sense, one could rename  these channels as 
``three-({\it visible})fermion'' processes. \\

Different aspects of  the four-fermion processes have been analyzed 
in \cite{lepd}. In the next section, 
we concentrate on the total cross sections
and present the rates for all the  four-fermion processes when 
some canonical cuts are imposed, as given by many available 
computer codes.

\noindent
Before doing that, we discuss the relevance of effective approximations in
studying four-fermions rates.
When including all the  tree-level diagrams  for a four-fermion  
process in a computer program, one can loose some insight
on   which subsets of  diagrams  are really dominant 
and which are ``sub-leading". On the other hand,  in order  
to treat correctly the phase-space integrations and to 
get a reliable result, one should 
distinguish the main/secondary groups of 
diagrams. At the same time, it is also useful to check the reliability
of effective approximations that allow to evaluate
given subsets of diagrams in a much simpler way.
The natural way of forming subsets of diagrams 
is by isolating subgraphs that
(with the in- and out- intermediate 
particles taken on mass shell) correspond to
some gauge-invariant process of lowest order. 

\noindent 
We illustrate such a procedure in the particular
process $e^+ e^- \rightarrow e^+ e^- b \bar b$. This channel  is 
important as a background for Higgs bosons searches.
Figure 11 shows the 48 diagrams 
 that make up the complete set (excluding the two that
involve Higgs bosons):
8 multiperipheral, 16  bremsstrahlung (single or non resonant,
with a $\gamma/Z$ in the $t$-channel),
8 conversion (single- or double-resonant) and
16 annihilation (single- or non-resonant) graphs.
The first three  classes of diagrams 
involve  the subprocesses
$\gamma \gamma \rightarrow b \bar b$, $\gamma e \rightarrow Ve$
($V=\gamma, Z$) and $e^+ e^- \rightarrow VV$, respectively. 
The contribution of each subset to the \underline{total} cross section
has been computed exactly at  tree level by the computer package 
CompHEP \cite{comp}, and then compared with the corresponding results 
obtained through  appropriate effective approximations.
These approximations involve a convolution of a simpler subprocess
of the four-fermion process considered
with either the equivalent photon spectrum in the Weizs\"acker-Williams
(WW) approximation or a conversion factor describing the decay 
of a virtual (or resonant) intermediate state.
Further details on these approximations can be found in \cite{lepd}.

\noindent
Note that, in general, the interferences between different subsets 
of graphs for $e^+ e^- \rightarrow e^+ e^- b \bar b$
are found to be negligible at  LEP2
energies, with the exception of the interferences
of the bremsstrahlung diagrams with the $Z\ra b \bar{b}$ decay,
and the conversion diagrams with the $\gamma^*\ra e^+e^-$ 
and $Z\ra b \bar{b}$ decays (that gives $-24$ fb at $\sqrt{s}=200$GeV).
Then, apart from the interference between the bremsstrahlung 
diagram
with $Z\ra b \bar{b}$ and the one with  $\gamma^*\ra b \bar{b}$,
which gives -3.2 fb at $\sqrt{s}=200$GeV, all other interferences 
are found to be less than 1 fb at the same energy \cite{dubi}.

\noindent
In figure 12, one can see that the exact computation
(solid) is always reasonably recovered by a proper approximation
(dashes). Indeed, adding the approximate formulae for multiperipheral,
single and double conversion  
incoherently (i.e., with no interferences)
the total cross section is reproduced within $5\%$.

\subsection{Cross sections for all four-fermion final states}
We now report on the results of a study of 
the tree-level cross sections for \underline{all} possible
four-fermion final states, 
as listed in  Tables~3-5. The complete set of diagrams  
is taken into account in each case (the corresponding 
total number of diagrams ($N_d$) is shown in the same tables).
Higgs-boson contributions are not included.
This comparative study involves seven computer codes:  
 ALPHA, CompHEP,
EXCALIBUR, grc4f (a package for computing 
four-fermion processes  based on GRACE),
WWGENPV/HIGGSPV, WPHACT
and WTO. For a detailed description of these codes  
see \cite{lepd}. 
In this comparison 
ISR and gluon-exchange diagrams for the hadronic four-fermion 
final states (when implemented) are switched off. 
The effect of non-zero fermion masses for some of the processes 
has also been investigated by ALPHA and grc4f
(see Tables~3-5). 
Total cross sections have been computed at the centre-of-mass
energy $\sqrt{s}=190$GeV, with the following cuts:
$E_{\ell^{\pm}}>1$GeV , $E_q>3$GeV , $\theta(\ell^{\pm}-beam)>10^o$ , 
$\theta(\ell^{\pm}-\ell'^{\pm})>5^o$ , $\theta(\ell^{\pm}-q)>5^o$ ,
$M_{qq^{(')}}>5$GeV (cuts on the fermion energy variables are 
loosened in the case of massive fermions).
Furthermore, in order to better check the agreement among 
the different
codes, a {\it canonical} set of input parameter has been
agreed upon in all the computations, that is 
$M_Z=91.1888$GeV, $\Gamma_Z=2.4974$GeV , $M_W=80.23$GeV ,
$\Gamma_W=\frac{3G_FM_W^3}{\sqrt{8}\pi}=2.0337$GeV , 
$\alpha^{-1}(2M_W)=128.07$, $G_F=1.16639\; 10^{-5}$GeV$^{-2}$ ,
$\sin^2 \theta_W$ from $\frac{\alpha(2M_W)}{2\sin^2 \theta_W}
=\frac{G_F M^2_W}{\pi \sqrt{2}}$. 
In Table~3, the cross sections for all the four-lepton final 
states are shown, in Table~4 the 
ones for the semileptonic states and in Table~5
the ones for the hadronic four-fermion states.
The error in the last one or two digits,
corresponding to the Monte Carlo event generator,
is also shown in parenthesis.
One can see that the agreement among the different central
values is  in general at the level of a few per-mil, and even better
in some cases. Note that, with the cuts above, the effect of the fermion 
masses can be not negligible, as can be seen, for instance,
by comparing the rates for 
muons to those for $\tau$'s ({\it cf.} Tables 3-4). 

\subsection{QCD tests}
LEP2 gives the possibility to test the QCD predictions 
for the $\epem$ physics at
an energy scale $Q \sim 180$ GeV, which is about twice the one at LEP1.
 Scaling violations are predicted to depend on $\log Q$ in QCD, 
which unfortunately corresponds to a rather modest variation of
the relevant phenomenology when going from LEP1 to LEP2.
Furthermore, the hadronic cross section at LEP2 is penalized by 
being far away from the $Z$ resonance. One has $\sigma_{had} \sim 20$
pb at LEP2 (i.e., about  3 order of magnitude less than at LEP1), 
that corresponds to a total of about $N \sim 10^4$ events 
(for 500 pb$^{-1}$), and  a typical statistical error of 
$1/\sqrt{N} \sim 1$\% on $\sigma_{had}$.
On the other hand, if one wants to check the evolution of the strong 
coupling constant $\alpha_s$, one of the theoretically  sounder method is
given by the measurement of the total hadronic cross section. 
Indeed, one has at the first order in $\alpha_s$
$$
\sigma_{had} \simeq \sigma_{had}^0(1+\frac{\alpha_s}{\pi}).
$$
Since at LEP2 energies QCD predicts $\alpha_s \simeq 0.107$
(on the basis of the LEP1 value and lower-energies determinations), 
this relation implies that a statistical error of 1\%
on $\sigma_{had}$ translates into an error of about 30\% on $\alpha_s$.
This error does not allow to appreciate experimentally the evolution
of $\alpha_s$ from the LEP1 energy ($\alpha_s(\mz)\simeq 0.122$)
down to  $\alpha_s(\simeq 180$GeV$)\simeq 0.107$, corresponding to a relative 
variation of only about 10\%.
Hence, the measurement of the hadronic cross section is not a useful 
method to determine $\alpha_s$ at LEP2.

The measurement of the jet fractions and different event-shapes variables
offers a  more sensitive procedure to determine $\alpha_s$,
which is nevertheless a bit less solid theoretically
due to the presence of non-negligible nonperturbative effects. 
The 3-jet fraction over a sample of 
about 10$^4$ hadronic events is expected to be of the order 10$^3$, 
depending directly on $\alpha_s$. This translates  into a statistical error on
$\alpha_s$ of about 3\%, that is sufficient to disentangle a 10\% evolution
from the LEP1 $\alpha_s$ value.
In figure~13 \cite{opce}, the preliminary results on the $\alpha_s$ measurement
at LEP2 energies by OPAL are reported along with lower-energy 
determinations and the next-to-leading order theoretical prediction
(continuous curves). Even with the present moderate statistics collected,
the QCD $\alpha_s$ evolution looks well reproduced by the LEP2 data.

Other QCD issues that can be explored at LEP2 are the fragmentation function
evolution and the charged multiplicity distributions. The fragmentation
functions can be well determined at LEP2, but,
theoretically,  little evolution is expected from LEP1.
On the other hand,  the charged multiplicity distributions
can test the available predictions of QCD in the leading-log approximation
combined with the local parton-hadron duality hypothesis.

\section{Discovery Physics}
LEP2 gives  the opportunity  to look for new particles
that interact only electroweakly (hence, hard to disentangle in 
hadronic machines), with masses in the range \\
$\mz/2 \lta m_{NEW} \lta 200$GeV, or even heavier in case they can be singly
produced.

In the standard model, the only (crucial) particle yet to be discovered
is the Higgs boson, that is needed within the 
framework of the spontaneous breaking of the electroweak symmetry.
This neutral (scalar) particle is in general hard to produce 
in ordinary collisions,
since its coupling with  other particles depends on the ratio of the particles
masses over the $W$ mass. Hence, the Higgs is very weakly coupled to the
electrons and quarks that interacts at high energies 
in a normal collider. In order to produce a Higgs with sufficient cross
section, one has first to excite some heavy degree of freedom, such as
$W$ and $Z$ bosons, from the initial beams. Then the Higgs can be produced,
either by radiation from a $Z/W$ or by fusion of a pair of
vector bosons. As a consequence, the typical cross section for Higgs
production al LEP does not exceed the 1-pb level, and is in general just a few
tenths of pb. This requires a sufficiently high integrated luminosity
[$\cal{O}(100^{-1})pb$] in order to eventually detect a reasonable signal.
Before the start of LEP2, the best direct limit on $\mh$ was found at LEP1,
where the range $\mh<64.5$GeV was excluded \cite{higl}.
Furthermore, by optimizing the fit to the standard model predictions
of all  the electroweak precision data at LEP1, one obtains an 
indirect upper bound on $\mh$ of about 300 GeV \cite{precd}.
On the other hand, LEP2 can approximately explore the $\mh$ range $\mh \lta
\sqs -100$GeV, and either discover a Higgs or improve considerably  the
LEP1 $\mh$ limits.

At LEP2, one can also explore possible extensions of the standard model 
predicting new particles with masses lighter than about 100 GeV.
A widely considered possibility is given by the models 
introducing SUperSYmmetry (SUSY) \cite {susy}.
A supersymmetric extension of the standard model offers 
the only way to make {\it natural} a theory with  fundamental 
scalar (Higgs) fields, such as the standard model. 
The Minimal Supersymmetric Standard Model (MSSM)
is the simplest, phenomenological model with low-energy SUSY 
effectively broken at an energy scale not much larger than $\mw$, 
as required by the solution of the naturalness problem.
It predicts the doubling of the particle spectrum, by associating
to each ordinary
particle  a new particle with spin that differs by 1/2 from the ordinary 
particle's spin, and  with same remaining quantum numbers. 
For instance, for each left- and right-handed fermionic
(quark's and lepton's) degree of freedom,  a scalar degree of freedom
with same quantum numbers is predicted [the squark  ($\tilde q$)
and the slepton ($\tilde \ell$)].
On the other hand, the partners of the spin-1 photon, gluon, W and Z
are the spin-1/2 photino, gluino, w-ino and z-ino.
Usually the mass-matrix  eigenstates mix the photinos and z-inos, on the 
one side,  and W-inos, on the other side, with the neutral and charged
components, respectively, of the higgsinos 
(the fermionic partners of the neutral and charged
higgs bosons, coming from the two Higgs doublets introduced in the MSSM).
Then, one obtains four neutral fermions [the neutralinos 
($\chi^0_i, i=1,\dots,4$)], and two charged fermions [the charginos
($\chi^{\pm}_i, i=1,\dots,2$)], which are predicted to be among the lightest
SUSY particles. In particular, the lightest neutralino ($\chi^0_1$)
could be the lightest SUSY particle. This has important phenomenological
consequences. Due to the conservation of the R parity, 
%**************************
\footnote{The R parity is a new multiplicative quantum number introduced
to avoid lepton and barion number violation in the SUSY lagrangian.
Its value is $+1$ for ordinary particles and $-1$ for SUSY partners.
If R parity is conserved, three important phenomenological consequences
follow: a) SuSy partners are produced in pairs in ordinary matter 
collisions. b) a SUSY particle always decays into an odd number of SUSY
particles. c) the lightest SUSY particle is stable.} 
%**************************
the neutralino is
then stable, and is present at the end of the decay chain of all the SUSY
particles.  Being also neutral and weakly interacting, it will show up 
inside a normal experimental apparatus just as a neutrino does, 
through some missing energy and momentum in the final
state. Hence, in general, the production of any SUSY particle
will be characterized by leptons and/or jets (and/or photons) accompanied
by missing energy and momentum.

Apart from the two main issues above,
LEP2 can also complement the TeVatron and HERA programs 
in the search for some signal from compositness and new interactions'
effects. For instance, after the 1996 runs, lower limits
of 1-2 TeV's \footnote{The exact value depends on the helicity features 
of the particular model considered.} on the energy scale
$\Lambda$ of a possible new contact interaction of the type $eeqq$ have been 
already put by OPAL \cite{opce}. Similarly, OPAL produced stronger limits on 
the mass of a possible new scalar particle (such as a leptoquark or a  squark)
exchanged in the $t$-channel in the process $e^+e^- \to q\bar q$.
These limits already improve the present bounds obtained at the
TeVatron and HERA. Excited leptons $\ell^*$ (with $\ell^*=e^*,\mu^*,\tau^*$)
are also being investigated, mainly through
the processes $e^+ e^- \to \ell \ell^* \to \ell^+ \ell^- \gamma$.

There are presently a few anomalous events at LEP1.5 and LEP2, 
that, if confirmed,
could be a signal of new physics. The ALEPH experiment observes
a clear excess in the four-jet production, that is kinematically compatible
with the production of two heavy particles (of mass  $m_1\sim 48$GeV 
and $m_2 \sim 58$GeV)
each decaying into two jets. The distribution of the dijet mass sum
(that is $\Sigma M=(m_1+m_2)$)
shows a clear peak around 106 GeV (see figure 14 \cite{alce}). In particular,
ALEPH observes 18 events in the interval $(106\pm 4)$GeV for $\Sigma M$,
against a background of 3.1 events expected.  
The problem is that all the other LEP experiments presently do not 
observe such a signal. 
This seems to point more to some experimental issue to solve  
than to some new phenomenon.

\subsection{Higgs search}
In figure~15, the two Feynman diagrams contributing to the Higgs boson
production at LEP2 are shown. The dominant contribution is given by
the process $e^+e^-\to H Z$, where the Higgs boson is radiated by a $Z$
vector boson (the so-called Higgs-strahlung process). 
On the other hand, the $WW$
fusion channel has smaller cross sections in general. It gives non-negligible
contributions only when the Higgs mass is beyond the threshold for
the Higgs-strahlung process, i.e., for $\sqs-\mz \lta \mh \lta \sqs$.
The total cross section for Higgs production versus $\mh$, for different
values of $\sqs$, is shown in figure~16 \cite{lepd}. 
In figure~17, the branching ratio BR for the main Higgs decay channels
is reported. One can see that the dominant decay is  the 
Higgs decay into a $b$ quark pair, with a corresponding BR$>$80\%.
Hence, the typical experimental signature for the $e^+e^-\to H Z$ channel 
is a pair of $b$-jets plus a jet or lepton pair coming from the $Z$ decay
(therefore, with invariant mass around $\mz$).
The four LEP experiment have carried out extensive simulations
in order to determine the exact potential of LEP2 for discovering the Higgs
boson.
The result of this study is summarized in figure~18 \cite{lepd}, 
where the minimum luminosity needed per experiment (in pb$^{-1}$) to 
discover  (solid line) and to exclude (dashed line) a Higgs boson is shown, 
as a function of $\mh$, for three values of $\sqs$.
On the other hand, in Table 6 \cite{lepd}, one can find the maximal $\mh$ 
that can be excluded or discovered with a given integrated luminosity 
per experiment.
All these results correspond to combining the data from the four 
experiments.
One can see that reaching 150 pb$^{-1}$ at $\sqs=192$GeV is sufficient
to cover the mass range up to 95 GeV for discovery and up to 98 GeV
for exclusion. Increasing further the luminosity improves
the situation only marginally. A possible run at $\sqs=205$Gev (not yet
approved at the moment), could exclude a Higgs as heavy as 112 GeV,
with a luminosity of 200 pb$^{-1}$.

After the 1996 LEP2 runs at $\sqs=161$ and 172 GeV, with a total 
luminosity of only
20 pb$^{-1}$, some improvement on the LEP1 limits on $\mh$ have
already been obtained. The best bound is the one by ALEPH
\cite{alce}, that gives $\mh>70.7$GeV. 

\subsection{Supersymmetry searches}
While  Higgs searches, due to the moderate production rates,  need
a rather large luminosity in order to exploit the full LEP2 potential,
the SUSY particles search is characterized by larger cross sections.
Hence, a modest luminosity can be sufficient
to either discover or exclude particle masses even near the kinematical limit.
In fact, one expects new {\it charged} particles in the lower SUSY mass 
spectrum, such as charginos and charged sleptons. Their electromagnetic
coupling guarantees cross sections of at least a few pb's.
This explains, for instance, the remarkable new LEP2 mass limit on charginos,
obtained from the short 1996 runs. Charginos are expected to be pair produced
at LEP via the reaction $e^+e^-\to \chi^+_1\chi^-_1$, whose cross section
depends on the physical composition of the chargino in terms of W-inos and
Higgsinos, and on the mass of the sneutrino that can be exchange in the
$t$-channel. In particular for relatively light sneutrino, the negative
interference between the $s$- and $t$- channel (mediated by a $\gamma/Z$ 
and a $\tilde \nu$, respectively) reduces the cross section, thus 
leading to worse limits.
The effects of the variation in the physical chargino composition 
and of the sneutrino mass
can be taken into account through a
scatter plot for the production rate versus the chargino mass
(see figure~19 \cite{dece}). 
In figure~19, the limits on $m_{\chi_1^+}$ found by DELPHI
in the 1996 run at 172 GeV for a heavy  (upper part) 
and a light (lower part) sneutrino are also shown.
For a heavy (light) sneutrino a lower bound of 84.5 (67.8) GeV  
at 95 \% of C.L. is found
(cf. the LEP1 limit on $m_{\chi_1^+}$ of  about 45 GeV). Also shown 
(dashed lines) are the bounds one obtains in alternative models,
where the lightest neutralino is not stable, but decays into a 
light gravitino plus a photon. 

By using unification conditions on the gaugino masses \cite{susy}, 
one can also extract a new limit on the lightest neutralino mass from 
the LEP2 chargino mass bound. For instance, figure~20 
\cite{alce} shows the ALEPH
bounds on the lightest neutralino mass from the data collected
at LEP1, LEP1.5 and LEP2, versus $\tan \beta$ (the ratio of the vacuum
expectation values for the two Higgs doublets, which usually parametrizes the
MSSM). At LEP1.5 and LEP2, one can finally
exclude a massless neutralino for {\it any} value of $\tan \beta$.
In particular, the mass range  $m_{\chi_1^0}<22$GeV  has been excluded 
by ALEPH at 95\% of C.L..

Finally, figure~21 \cite{alce} presents the ALEPH 1996 
limits on the charged slepton 
masses (separately, for each flavour, and combined). 
The combined bound is
about 75 GeV, that is quite lower than the chargino bound.
This is because of the threshold factor in the cross section 
for scalar pair production ($\sim \beta^3$,
being $\beta$ the velocity of the produced scalars)
that reduces the rates near the kinematic limit, with respect
to the fermion pair  production.

Another crucial part of the MSSM to test at LEP2 is the Higgs sector.
In the MSSM, 
the lightest neutral higgs corresponding to the two 
doublets (which gives rise to two neutral scalar, one pseudoscalar and 
one charged Higgs bosons) should be lighter than the 
standard model higgs boson, and  hence more likely to be produced
at LEP2. An extensive analysis has been done in this field.
The two main production channels are the Higgs-strahlung and the associated
production of the lightest neutral scalar $h$ and the pseudoscalar $A$.
Table~7 \cite{lepd} (analogously to Table~6) 
summarizes the expectations for the 
discovery and exclusion of SUSY Higgs bosons at LEP2. 

\section{Conclusions}
The second phase of LEP just started, and produced 
the first results both in the field of the standard model tests and on 
new physics bounds.
Further crucial progresses are expected from the full machine operation in the
years 1997-1999, especially in the $W$ boson physics and in the 
exploration of the Higgs sector of the standard model.
With an optimistic view, one could wait for the discovery
of new particles/interactions effects, that are indeed possible
with the collision energy available at LEP2.
A more conservative attitude is anyway assured of new results
in the precision measurement of the standard model and in  constraining 
possible new theories, that will nicely complement the experimental analysis 
from the TeVatron and HERA.

\vspace{0.7cm} 
\noindent 
{\large \bf Acknowledgements} 

\noindent
The author wishes to thank all the organizers,
and in particular J.A. Helay\"el Neto and Gentil O. Pires,  for the warm
hospitality and the excellent organization of the School. 

\end{document}